\documentclass[conference]{IEEEtran}%conference,a4paper,

\IEEEoverridecommandlockouts

\usepackage{textcomp}
\usepackage[pdftex]{graphicx}
\usepackage{marginnote}
\usepackage{multirow}
\usepackage{cite}
\usepackage{booktabs,threeparttable}
\usepackage{etoolbox}
\appto\TPTnoteSettings{\footnotesize}
\usepackage[strict]{csquotes}
\usepackage{amsmath, amsthm, amssymb, amsfonts, array,soul,color,bm, mathtools}
\usepackage{xcolor}

\usepackage{balance}
\usepackage{subcaption}

\usepackage{caption}
\usepackage{color}
\usepackage{tikz} % for tikz
\usepackage{pgfplots} 
\usepackage{pgfgantt}
\usepackage{pdflscape}

\usepackage[linesnumbered,ruled,vlined]{algorithm2e}
\usepackage{algpseudocode}

\DeclareMathOperator*{\argmin}{arg\, min}

\usepackage{siunitx}
\usepackage{url}
\usepackage{marginnote}

\usepackage{array}
\newcolumntype{P}[1]{>{\centering\arraybackslash}p{#1}}
\newcolumntype{U}[1]{>{\centering\arraybackslash}u{#1}}

\usepackage{microtype} 
\usepackage{relsize}
\usepackage{glossaries}
\newacronym{3gpp}{3GPP}{the 3rd generation partnership project}
\newacronym{a-ag}{A-agent}{anchor-agent}
\newacronym{ap}{AP}{access point}
\newacronym{aoa}{AoA}{angle of arrival}
\newacronym{aod}{AoD}{angle of departure}
\newacronym{bs}{BS}{base station}
\newacronym{ru}{RU}{radio unit}
\newacronym{brsrp}{B-RSRP}{beam reference signal received power}
\newacronym{crlb}{CRLB}{Cram\'{e}r-Rao lower bound}
\newacronym{cfp}{CFP}{convex feasibility problem}
\newacronym{d2d}{D2D}{device-to-device}
\newacronym{ekf}{EKF}{extended Kalman filter}
\newacronym{fim}{FIM}{Fisher information matrix}
\newacronym{gnb}{gNB}{new generation node B}
\newacronym{gn}{GN}{Gauss-Newton}
\newacronym{iot}{IoT}{Internet of things}
\newacronym{iiot}{IIoT}{industrial Internet of things}
\newacronym{ldm}{LRMs}{location-related measurements}
\newacronym{los}{LoS}{line-of-sight}
\newacronym{lte}{LTE}{long-term evolution}
\newacronym{jrcup}{JrCUP}{joint RIS calibration and user positioning}
\newacronym{simo}{SISO}{single-input multiple-output}
\newacronym{mtc}{MTC}{machine type communications}
\newacronym{mmw}{mmWave}{millimeter wave}
\newacronym{mimo}{MIMO}{multiple-input multiple-output}
\newacronym{nlos}{NLoS}{non-line-of-sight}
\newacronym{nr}{NR}{New Radio}
\newacronym{noma}{NOMA}{non-orthogonal multiple access}
\newacronym{ofdm}{OFDM}{orthogonal frequency division multiplex}
\newacronym{ofdma}{OFDMA}{orthogonal frequency division multiple access}
\newacronym{prose}{ProSe}{proximity-based services}
\newacronym{pscch}{PSCCH}{physical sidelink control channel}
\newacronym{pssch}{PSSCH}{physical sidelink shared channel}
\newacronym{qos}{QoS}{quality of service}
\newacronym{ris}{RIS}{reconfigurable intelligent surface}
\newacronym{rss}{RSS}{received signal strength}
\newacronym{rwp}{RWP}{random waypoint}
\newacronym{scns}{SCNs}{small cell networks}
\newacronym{sdp}{SDP}{semi-definite programming}
\newacronym{siso}{SISO}{single-input single-output}
\newacronym{snr}{SNR}{signal-to-noise ratio}
\newacronym{sops}{CPS}{cooperative positioning system}
\newacronym{slam}{SLAT}{simultaneous localization and tracking}
\newacronym{srs}{SRS}{sounding reference signal}
\newacronym{mt-sops}{MT-SOPS}{multi-target self-organized positioning system}
\newacronym{tdoa}{TDoA}{time difference of arrival}
\newacronym{toa}{ToA}{time of arrival}
\newacronym{t-ag}{T-agent}{target-agent}
\newacronym{v2v}{V2V}{vehicle-to-vehicle}
\newacronym{v2x}{V2X}{vehicle-to-everything}
\newacronym{ue}{UE}{user equipment}
\newacronym{ul}{UL}{uplink}
\newacronym{ura}{URA}{uniform rectangle array}
\newacronym{ula}{ULA}{uniform linear array}
\newacronym{uwb}{UWB}{ultra-wideband}
\newacronym{wrt}{w.r.t.}{with respect to}
\newacronym{wcg}{WCG}{weighted centroid geometry}

\long\def\comment#1{}

\newfont{\bbb}{msbm10 scaled 700}

\newcommand{\hthickline}{\noalign{\hrule height 0.80pt}}

\newfont{\bb}{msbm10 scaled 1100}

\newcommand{\av}{{\bf a}}

\newcommand{\hv}{{\bf h}}

\newcommand{\nv}{{\bf n}}
\newcommand{\ov}{{\bf o}}
\newcommand{\pv}{{\bf p}}

\newcommand{\sv}{{\bf s}}
\newcommand{\tv}{{\bf t}}

\newcommand{\wv}{{\bf w}}

\newcommand{\Hm}{{\bf H}}

\newcommand{\Jm}{{\bf J}}

\newcommand{\Rm}{{\bf R}}

\newcommand{\Gc}{{\cal G}}

\newcommand{\etav}{\hbox{\boldmath$\eta$}}

\newcommand{\varphiv}{\hbox{\boldmath$\varphi$}}

\newcommand{\Omegam}{\hbox{\boldmath$\Omega$}}

\newcommand{\diag}{{\hbox{diag}}}

\def\BibTeX{{\rm B\kern-.05em{\sc i\kern-.025em b}\kern-.08em
    T\kern-.1667em\lower.7ex\hbox{E}\kern-.125emX}}

\begin{document}

\title{Joint RIS Calibration and Multi-User Positioning}%towards Intelligent 6G Wireless

\author{Yi Lu\IEEEauthorrefmark{1}, Hui Chen\IEEEauthorrefmark{2}, Jukka Talvitie\IEEEauthorrefmark{1}, Henk Wymeersch\IEEEauthorrefmark{2}, and Mikko Valkama\IEEEauthorrefmark{1}\\
\IEEEauthorrefmark{1}Electrical Engineering, Tampere University, Finland\\
\IEEEauthorrefmark{2}Chalmers University of Technology, Sweden\\
e-mail: yi.lu@tuni.fi, hui.chen@chalmers.se
}

% make the title area
%\maketitle
{\let\newpage\relax\maketitle}
{\let\thefootnote\relax\footnotetext{\footnotesize
This work was financially supported by the Academy of Finland under grants \#319994, \#338224, \#328214, \#323244,  and through the H2020 project Hexa-X (Grant Agreement no. 101015956). }}
%Online videos are available at: \textcolor{blue}{\url{}}

\begin{abstract}
Reconfigurable intelligent surfaces (RISs) are expected to be a key component enabling the mobile network evolution towards a flexible and intelligent 6G wireless platform. In most of the research works so far, RIS has been treated as a passive \gls{bs} with a known state, in terms of its location and orientation,  to boost the communication and/or terminal positioning performance. However, such performance gains cannot be guaranteed anymore when the RIS state is not perfectly known. In this paper, by taking the RIS state uncertainty into account, we formulate and study the performance of a joint RIS calibration and user positioning (JrCUP) scheme. From the Fisher information perspective, we formulate the JrCUP problem in a network-centric single-input multiple-output (SIMO) scenario with a single BS, and derive the analytical lower bound for the states of both user and RIS. We also demonstrate the geometric impact of different user locations on the JrCUP performance while also characterizing the performance under different RIS sizes. Finally, the study is extended to a multi-user scenario, shown to further improve the state estimation performance.
\end{abstract}
%\\ \textcolor{red}{We will re-fine the Abstract in the end.}
\begin{IEEEkeywords}
5G New Radio, 6G, Fisher information, joint calibration and positioning,  reconfigurable intelligent surfaces
\end{IEEEkeywords} 

\IEEEpeerreviewmaketitle

% positioning or localization, choose one and unify throughout this paper
\section{Introduction}

Evolving from 5G to 6G, the wireless networks are transforming into a ubiquitous, intelligent and multi-function service platform with the support of several key technical enablers, such as artificial intelligence, cognitive slicing, proactive channel coding, and \gls{ris}~\cite{THz_Sensing_Loca_2020,RISE_6G_2021}. Among them, the \gls{ris} technology is seen as a promising transformative component to realize smart connectivity, which also ties the three fundamental wireless applications, i.e., communications, localization and sensing/mapping  together~\cite{Loca_Map_RIS_VTM_2020}. Besides the potential benefits in communications~\cite{Wireless_Through_RIS_2019,BMIMO_LIS_2018,RIS_Emil_2020,Comm_W_LIS_2020}, \gls{ris} can also be effectively deployed to construct a controllable and reconfigurable channel for improved positioning performance with lower costs than, e.g., the ultra-dense deployments of 5G \gls{nr} \gls{bs}s~\cite{book_UDN_2020}.

% a reflection (multipath)-assisted wireless system for efficient positioning. That being said, instead of passively decomposing and processing different path components, the active utilization and deployment of \gls{ris} help form a controllable and reconfigurable wireless environment for improved positioning performance with lower cost than the ultra-dense deployment of \gls{nr} \gls{bs}s~\cite{book_UDN_2020}, for instance. 

Earlier works on \gls{ris}-enabled positioning include~\cite{FPrint_RIS_2021,ANN_RIS_Loca_2021, B5G_RIS_LoS_2020, RIS_enabled_UserPos_2021,SISO_RIS_LoS_2021,AdaBF_4_RISaid_Joint_Comm_Loca_2020, PAPIR_RIS_Loca_reflection_2021, NLoS_RIS_2021,Near_Field_RIS_loca_2022,RIS_PEB_OEB_2021,Indoor_RIS_2021,Indoor_RIS_SLAM_2021,chen2021tutorial}, covering fingerprinting  approaches as well as methods based on geometric channel parameters (i.e., angles and delays). 
For fingerprinting-based approaches, the utilization of \gls{ris} has provided reduced computational complexity and improved accuracy, as shown in~\cite{FPrint_RIS_2021,ANN_RIS_Loca_2021}. By employing the delay and angle measurements obtained from radio signals, the positioning performance with \gls{ris}s as reflectors or scatters has been studied in different situations, e.g., in the \gls{los} (with respect to the \gls{bs})~\cite{B5G_RIS_LoS_2020, RIS_enabled_UserPos_2021,SISO_RIS_LoS_2021}, the corresponding \gls{nlos}~\cite{AdaBF_4_RISaid_Joint_Comm_Loca_2020, PAPIR_RIS_Loca_reflection_2021, NLoS_RIS_2021,Near_Field_RIS_loca_2022}, in the downlink side~\cite{B5G_RIS_LoS_2020} or the uplink side~\cite{PAPIR_RIS_Loca_reflection_2021,RIS_PEB_OEB_2021}, and in indoor environments~\cite{Indoor_RIS_2021,Indoor_RIS_SLAM_2021}. Finally, 
%In particular, the \gls{siso} 2D scenario where both the \gls{bs} and the user equip a single antenna has been studied in~\cite{B5G_RIS_LoS_2020,SISO_RIS_LoS_2021,RIS_enabled_UserPos_2021} from \gls{fim} perspective together with the design of \gls{ris} phase profile. Moving to the \gls{mimo} configuration, the authors of~\cite{RIS_PEB_OEB_2021,AdaBF_4_RISaid_Joint_Comm_Loca_2020} have investigated a \gls{ris}-aided millimeter wave system under \gls{los} and \gls{nlos} conditions, and
a comparison between 5G and 6G systems is provided in~\cite{chen2021tutorial}. %Both works have highlighted the importance of the design of the \gls{ris} phase profile to the achievable performance. Compared to the situation where the \gls{ris}s are deployed separately with respect to the user, a \gls{ris}-equipped positioning scenario with multiple \gls{bs}s was considered in~\cite{RIS_enabled_UserPos_2021}. A batch of delay measurements was estimated and applied to perform user positioning. Even under the \gls{nlos} condition where the direct path between the \gls{bs} and user are obstructed~\cite{AdaBF_4_RISaid_Joint_Comm_Loca_2020, PAPIR_RIS_Loca_reflection_2021, NLoS_RIS_2021,Near_Field_RIS_loca_2022}, with well placed \gls{ris}, useful channel information can be extracted from the \gls{ris}-relayed path for positioning and / or communication functions. 

% fig. 1, system illustration for problem formulation
\begin{figure}[!t]% figure
\centering
\includegraphics[width=0.99\linewidth]{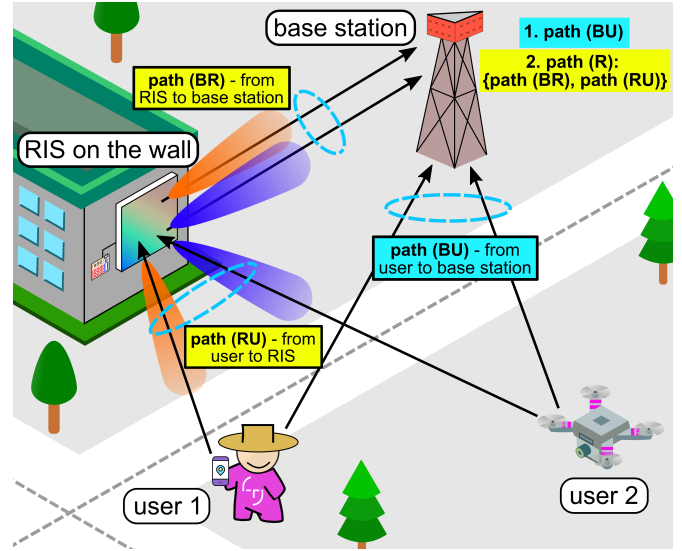}
\caption{A graphical illustration of the considered \gls{ris}-integrated network for joint \gls{ris} calibration and user positioning (JrCUP). Direct user-BS paths are marked in cyan, while paths including the \gls{ris} are marked in yellow. %The \gls{ris}-relayed path, i.e., path (R) includes essentially two sub-paths, path (RU) (from user to \gls{ris}) and path (BR) (from \gls{ris} to \gls{bs}) 
}\label{fig:RIS_sys}
\end{figure}
%\vspace{-2mm}

An important limitation in all the above works is that the 
%So far, most of the existing works have treated the 
\gls{ris} location and orientation %is as available prior information, i.e., the state of \gls{ris}, including the location and orientation 
were assumed to be precisely known. With such assumption, \gls{ris}s can therefore be treated as secondary \gls{bs}s or anchor points to improve the positioning and/or communication performance. However, in reality, due to potential deployment faults, external disturbances, and/or improper installation, the state of \gls{ris} may not be perfectly known, thus, needing further calibration. Towards this end, we formulate and investigate the feasibility and performance of a \gls{jrcup} scheme in this paper, where the state of \gls{ris} is not assumed to be perfectly known. The contributions of this work are as follows:
% The achievable accuracy of both users and \gls{ris} are presented. Specifically,
\begin{itemize}
    \item We formulate the \gls{jrcup} problem in the 3D scenario, where the states of both users and the RIS need to be jointly estimated;
    \item We carry out the \gls{fim} analysis, and compute the lower bound of the estimated channel parameters and states of user and \gls{ris}, while also address their dependence on the problem geometry;
    \item We propose a low-cost 2D searching-based initialization method and devise an efficient snapshot-like joint state estimation method in the general multi-user scenario; 
    \item We provide an extensive set of numerical results, showing that the lower bound can be achieved by the proposed estimation method, while also demonstrating that the multi-user scenario provides improved state estimation performance compared to the single-user case. 
\end{itemize}

%This paper is organized as follows. The applied system geometry model, signal model and RIS phase profile are outlined in Section~\ref{sec:system_Descrip}. The analytical performance bound together with the initialization and estimation algorithm are given in Section~\ref{sec:Bound_and_Algorithm}, followed by Section~\ref{sec:numerical_results} where the numerical results are presented. Finally, the conclusions and potential future directions are drawn in Section~\ref{sec:conclusion}.

%\red{will come back to write more details.}

% the near-field~\cite{RIS_PEB_OEB_2021,Near_Field_RIS_loca_2022}, the far-field~\cite{AdaBF_4_RISaid_Joint_Comm_Loca_2020,PAPIR_RIS_Loca_reflection_2021,RIS_enabled_UserPos_2021},
%In this paper, we investigate a scenario where the state of \gls{ris} is nor precisely known, while localizing the user location, i.e., a joint user localization and \gls{ris} calibration scheme.

\iffalse
Finally, the main contributions of the proposed positioning framework in this paper are summarized as:
\begin{enumerate}
    \item By doing A and B, we ;
    \item By fusing C and D, the algorithm;
    \item With ..., we evaluate the impact;
    \item Last but not least, we investigate . 
\end{enumerate}

\fi

\section{System Model}
\label{sec:system_Descrip}

In this work, a 3D outdoor positioning scenario is considered and illustrated in Fig.~\ref{fig:RIS_sys}, where there exists several users, one \gls{ris} deployed on the  surfaces of a building, and a \gls{bs}. The state of \gls{bs} is known and used as the reference point in the considered coordinate system, while the states of users and \gls{ris} are to be estimated. Facilitating a network-centric wireless system, the \gls{bs} observes and processes the uplink signals received from indirect paths via RIS and the direct path from user(s) to BS. 
%two paths: one from the user(s) directly which we denote as path (BU) and marked in cyan, and the other that starts from the user(s), through the \gls{ris}, and finally received at the \gls{bs} which we denote as path (R) and marked in yellow. The \gls{ris}-relayed path, i.e., path (R) includes essentially two sub-paths, path (RU) (from user to \gls{ris}) and path (BR) (from \gls{ris} to \gls{bs}).
% For the sake of clear notation, the direct path(s) from the user(s) to the base station is denoted as path (0), whereas the \gls{ris}-relayed path(s) from the user(s) to base station is denoted as path (1+2) 

\subsection{Geometry Model}\label{subsec:Geometry}

Without loss of generality, we start the system descriptions by considering one arbitrary user and the \gls{ris}, while the extension towards multi-user and multi-\gls{ris} will be treated later. Specifically, a single-antenna user, an $N_\text{R}$-element RIS and an $N_\text{B}$-antenna BS located at $\pv_\text{U} = [x_{\text{U}}, y_{\text{U}}, z_{\text{U}}]^\top$, $\pv_\text{R} = [x_{\text{R}}, y_{\text{R}}, z_{\text{R}}]^\top$ and $\pv_\text{B} = [x_{\text{B}}, y_{\text{B}}, z_{\text{B}}]^\top$, respectively, are considered where $\pv_\text{B} $ is the center of the BS array defined as the origin of the coordinate system (the BS is facing x-axis by default).
The direction vector from the BS to the user can be expressed as
\begin{equation}
    \tv_\text{BU} = -\tv_\text{UB} =
    \begin{bmatrix}
    t_{\text{BU}, x}\\
    t_{\text{BU}, y}\\
    t_{\text{BU}, z}
    \end{bmatrix}
    = \frac{ \pv_\text{U}- \pv_\text{B}}{d_\text{BU}},
    %= \frac{ \pv_\text{U}- \pv_\text{B}}{\Vert \pv_\text{U}-\pv_\text{B} \Vert},
    \label{eq:global_direction_vector}
\end{equation}
where %$\Vert \cdot \Vert$ is denoted as the Euclidean distance of the vector, and 
$d_\text{BU}=\Vert \pv_\text{U}-\pv_\text{B} \Vert$ is the distance between BS and user. Similar definitions apply to the RIS-user (RU) and BS-RIS (BR) paths.
Furthermore, the orientation of RIS is defined as a 3D Euler angle vector $\ov_\text{R} = [o_1, o_2, o_3]^\top$ (pitch, roll, yaw) while the transformation from Euler angle vector $\ov_\text{R}$ to the rotation matrix $\Rm_\text{R}$ can be found in~\cite{Rot_Mat_2019}.% [ref].

In terms of the \gls{aoa} and \gls{aod}, we denote $\varphiv_{\text{BU}}$ as the \gls{aoa} from user to \gls{bs}, $\varphiv_{\text{RU}}$ as the \gls{aoa} from user to \gls{ris}, $\varphiv_{\text{RB}}$ as the \gls{aod} from \gls{ris} to \gls{bs}, and $\varphiv_{\text{BR}}$ as the \gls{aoa} from \gls{ris} to \gls{bs}, where each $\varphiv = [\phi, \theta]^\top$ consists of azimuth and elevation angle and is measured in the local coordinate system. Since the BS is assumed to be the coordinate origin, its local and global direction vectors are identical. %Moreover, single antenna users are considered in this work, therefore, the angle-related measurements will not be available at the user side. 
For the sake of clarity, we define the local direction vector $\tilde \tv$ of the channel from \gls{bs} to RIS as 
\begin{equation}
    \tilde \tv_{\text{RB}}\left(\bm{\varphi}_{\text{RB}}\right) = \Rm^{\top}_\text{R}\tv = 
    \begin{bmatrix}
    \cos( \phi_{\mathrm{RB}})\cos( \theta_{\mathrm{RB}}) \\
    \sin( \phi_{\mathrm{RB}})\cos( \theta_{\mathrm{RB}}) \\
    \sin( \theta_{\mathrm{RB}})
    \end{bmatrix}.
    \label{eq:dir_vec_from_angle}
\end{equation}

\subsection{Signal Model}\label{subsec:sig_mod}
We consider an OFDM-based uplink communication system, where two radio paths are observed and processed at the BS. 
The received frequency domain signal vector at the $g$th transmission (one transmission means one OFDM symbol transmission) and $k$th subcarrier can be formulated as~\cite{chen2021tutorial}
\begin{equation}\label{eq:y_gk}
\begin{aligned}
    y_{g,k} &= \wv_{\text{B},g}^\top \left[\hv_{g,k} x_{g,k} + \nv_{g,k}\right], % \\ 
    %&= \wv_{\text{B},g}^\top \left[(\Hm_\text{BU} + \Hm_{\text{R},g}) x_{g,k} + \nv_{g,k}\right],
\end{aligned}
\end{equation}
where $\wv_{\text{B},g}\in \mathbb{C}^{N_\text{B}\times1}$ is the combiner vector at the BS, and $\nv_{g,k} \in \mathbb{C}^{N_\text{B}\times1}$ refers to the additive white Gaussian noise vector with a complex normal distribution $\mathcal{CN}(0, \sigma^{2}\bm{I}_{N_{B}})$. The channel $\Hm_{g,k}$ comprises two parts: the direct channel from user to BS and the reflected channel via the RIS, i.e., $\hv_{g,k}=\hv_{\text{BU},k} + \hv_{\text{R},g,k}$. 

The channel from user to BS is given by 
\begin{align}
    \hv_{\text{BU},k} = \alpha_\text{BU} {\av_\text{BU}}(\varphiv_{\text{BU}})e^{-j 2 \pi \Delta_f k \tau_{\text{BU}}}
    %= \frac{\lambda_c}{4\pi d_\text{BU}}e^{-j \frac{2\pi}{\lambda_c} d_\text{BU}}\av_\text{BU},
\end{align}
where $\Delta_f$ is the subcarrier spacing, $\tau_{\text{BU}}=d_{\text{BU}}/c + \beta$ represents the delay of BU path (including the synchronization offset $\beta$). The speed of light is denoted as $c$. Additionally, $\alpha_\text{BU}=\rho_\text{BU}+j\xi_\text{BU}$ is the complex gain of the LOS channel, $\av_\text{BU}(\varphiv_{\text{BU}})$ is the steering vector of the BS-user channel with the $b$th element ($1\le b\le N_\text{B}$), expressed as
\begin{align}
    \av_{\text{BU},b}(\varphiv_{\text{BU}}) = e^{j\frac{2\pi} {\lambda_c} \pv_b^\top \tilde\tv_\text{BU}},
    \label{eq:steering_vector_BU}
\end{align}
and $\pv_b$ is the local position of the $b$th element with respect to the array center. We note that $\pv_{b, x}$ is equal to \textbf{0} since a uniform planar array (UPA) is assumed.

%: 1) a user to BS \gls{los} channel $\Hm_\text{BU}$ and 2) a RIS-relayed channel $\Hm_\text{R}$. The UE is assumed to have a synchronization offset $\beta$ with respect to the clock offset of the \gls{bs}.  In particular, the user to BS \gls{los} channel $\Hm_\text{BU}$ is defined as
%\begin{align}
 %   \Hm_\text{BU} &= \alpha_\text{L} e^{-j \frac{2\pi}{\lambda_c} d_\text{BU}}{\av_\text{BU}} 
  %  = \frac{\lambda_c}{4\pi d_\text{BU}}e^{-j \frac{2\pi}{\lambda_c} d_\text{BU}}\av_\text{BU},
%\end{align}
%where $\alpha_\text{BU}=\rho_\text{BU}+j\xi_\text{BU}$ is the complex gain of the LOS channel, $\av_\text{BU}$ is the steering vector of the BS-UE channel with the $b$th element ($1\le b\le N_\text{B}$) as
%\begin{align}
 %   \av_{\text{BU},b} = e^{j2\pi f_c \pv_b^\top \tilde\tv_\text{BU}},
 %   \label{eq:steering_vector_BU}
%\end{align}
%and $\pv_b$ is the local position of the $b$th element with respect to the array center (Note that $\pv_{b, x}$ equals to 0 since a UPA is assumed). 
The RIS-relayed channel $\hv_{\text{R},g,k}$ is defined as
\begin{align}
    \hv_{\text{R},g,k} &= \alpha_\text{R} {\av_\text{BR}}(\varphiv_{\text{BR}})({\av_\text{RB}}(\varphiv_{\text{RB}}))^\top\Omegam_g \av_\text{RU}(\varphiv_{\text{RU}})e^{-j 2 \pi \Delta_f k \tau_{\text{R}}}%\\
    %& = \frac{\lambda_c^2}{16\pi^2 d_\text{BR}d_\text{RU}} {\av_\text{BU}}\text{diag}(\Omegam_g)^\top(\av_\text{RB}\odot\av_\text{RU}),
\end{align}
where $\alpha_\text{R}=\rho_\text{R}+j\xi_\text{R}$ is the complex channel gain of the RIS path, 
$\tau_{\text{R}}=(d_{\text{BR}}+d_{\text{RU}})/c+\beta$ refers to the delay of the path R, and $\Omegam_g = \diag{[\omega_{1,g}, \ldots, \omega_{N_\text{R},g}]}$ is the time-varying RIS configuration, with $|\omega_{i,g}|=1$, $\forall i,g$. Furthermore, $\av_\text{RB}(\varphiv_{\text{RB}})$ and $\av_\text{RU}(\varphiv_{\text{RU}})$ are the steering vectors that can be obtained from~\eqref{eq:steering_vector_BU}. We further define an intermediate steering vector at the RIS as $\av_\text{R}(\varphiv_{\text{RB}},\varphiv_{\text{RU}}) = \av_\text{RB}(\varphiv_{\text{RB}})\odot\av_\text{RU}(\varphiv_{\text{RU}})$ and its $r$th element can be obtained as
\begin{equation}
    \av_{\text{R},r}(\varphiv_{\text{RB}},\varphiv_{\text{RU}}) = e^{j\frac{2\pi} {\lambda_c} \pv_r^\top \tilde\tv_\text{R}} = e^{j\frac{2\pi} {\lambda_c} \pv_r^\top (\tilde\tv_\text{RB} + \tilde\tv_\text{RU})},
    \label{eq:steering_vec_RIS_intermediate}
\end{equation}
where $\tilde\tv_\text{R} = [\vartheta_1, \vartheta_2, \vartheta_3]^\top$ is the intermediate direction vector. 

\textbf{\textit{Remark:}} Considering that the first entry of the $\pv_b$ is 0 and the fact that RIS does not perform any signal processing, the following intermediate \gls{aoa} measurements can be obtained:
\begin{equation}\label{eq:coupled_AoA}
\begin{aligned}
& \vartheta_2 = \sin(\phi_\text{RU})\cos(\theta_\text{RU}) + \sin(\phi_\text{RB})\cos(\theta_\text{RB}) \\
& \vartheta_3 = \sin(\theta_\text{RU}) + \sin(\theta_\text{RB}).
\end{aligned}
\end{equation}
In other words, there are overall four angle measurements ($\phi_\text{RB}, \theta_\text{RB}, \phi_\text{RU}, \theta_\text{RU}$) involved and correlated with the RIS-relayed channel. However, only two intermediate angles~\eqref{eq:coupled_AoA} can be acquired and estimated, resulting in 8 measurements. The fundamental purpose of this work is to estimate the RIS location, the RIS orientation, the user location and the clock offset simultaneously, based on these 8 measurements.
%containing RIS-related angles

\subsection{RIS Profile and BS Combiner Design}\label{subsec:RIS_profile}
In this work, we assume that no prior information about the RIS and UE states is available. In such a case, we use random RIS profiles for each transmission, i.e., {the coefficient of the $i$th RIS element in the $g$th transmission, $\omega_{i,g}$, is assigned with a unit amplitude $|\omega_{i,g}|=1$ and a random phase following a uniform distribution as $\angle \omega_{i,g} \sim\mathcal{U}(-\pi,\pi)$. Similar random coefficients are also considered for BS combiner vectors across different transmissions. Such a method does not require any prior information of the RIS or the user(s), nor any specifically designed codebook, but necessitates a large number of transmissions, thus yielding increased latency.} The optimization of the RIS profile and the BS combiner with prior information (e.g., iteratively obtained with localization feedback) will be considered in our future work.

\section{Performance Bound and Estimation Algorithm}
\label{sec:Bound_and_Algorithm}

In this section, we outline and derive the lower bound of the parameters of interest using the ubiquitous \gls{fim}. Thereafter, we present the proposed initialization and snapshot estimation methods for \gls{jrcup}. %joint user positioning and \gls{ris} calibration.

\subsection{From FIM to Lower Bound}\label{eq:FIM}%Fisher Information Matrix

In general, the analytical lower bound of the parameters of interest can be obtained by computing the corresponding \gls{fim} based on the observed measurements. Herein, the observed measurements are the received signals at the \gls{bs}, obtained by collecting~\eqref{eq:y_gk} over the considered transmissions and active subcarriers, yielding $\bm{Y}\in\mathbb{C}^{G\times K}$ where $\bm{Y}=[\bm{y}_{1},\cdots,\bm{y}_{G}]$ and $\bm{y}_{g}\in\mathbb{C}^{K\times 1}$, in which $G$ is the total number of OFDM transmission and $K$ refers to the number of subcarriers. Thereafter, the \gls{fim} of the channel parameters can be computed as follows~\cite{Kay_1993}
\begin{equation}\label{eq:FIM_channelPara}
    \mathbf{I}(\etav) = \frac{2}{\sigma^2}\sum^{\Gc}_{g=1} \sum^K_{k=1}\mathrm{Re}\left\{
    \left(\frac{\partial\mu_{g,k}}{\partial\etav}\right)^H 
    \left(\frac{\partial\mu_{g,k}}{\partial\etav}\right)\right\}
    \end{equation}
where $\mu_{g,k}=\wv_{\text{B},g}^\top \hv_{g,k} x_{g,k}$ is the noise-free version of the observed symbol $y_{g,k}$ in~\eqref{eq:y_gk}. Moreover, the channel parameter vector is denoted as
\begin{equation}\label{eq:meas_vec}
 \etav=[\varphiv_\text{BU}, \varphiv_\text{BR}, \vartheta_{2}, \vartheta_{3}, \tau_\text{BU}, \tau_\text{R}, \rho_\text{BU}, \rho_\text{R}, \xi_\text{BU}, \xi_\text{R}]^{\top}   
\end{equation}
where the last four are the nuisance parameters, in which $\rho$ and $\xi$ refer to the real and imaginary parts of the channel gain. The sub-indices $(\cdot)_\text{BU}$ and $(\cdot)_\text{R}$ refer to the gains of the  user to BS (BU) and the \gls{ris}-relayed (R) paths, respectively. More importantly, the first 8 parameters\footnote{Note that $\bm{\varphiv}=[\phi,\theta]^{\top}$ defined in Sec.~\ref{subsec:Geometry} contains both azimuth and elevation angles, therefore, the overall number of available measurements is 8.} are the geometry-related measurements from which we extract the information of both the user and the RIS states.

By performing Schur complement~\cite{SchurCompl_2005}, an effective \gls{fim} of~\eqref{eq:FIM_channelPara} can be obtained with the nuisance parameters (i.e., the channel gains) being removed. Thereafter, the \gls{fim} of the state parameter vector can be calculated as
\begin{equation}\label{eq:FIM_statePara}
    \mathbf{I}(\sv) = \Jm_\mathrm{S}^\top \mathbf{\acute{I}}(\etav) \Jm_\mathrm{S},
\end{equation}
where $\mathbf{\acute{I}}(\etav)$ is the effective \gls{fim} computed from~\eqref{eq:FIM_channelPara}, and $\Jm_\mathrm{S}\triangleq \partial \etav/\partial \sv$ represents the Jacobian matrix, essentially the derivative of the channel parameters with respect to the state parameters. In particular, the state parameter vector $\sv\in\mathbb{R}^{8\times 1}$ is defined as
\begin{equation}\label{eq:State_Vec}
    \sv = [\sv_{\text{R}}^{\top}, \sv_{\text{U}}^{\top}]^{\top} = [\pv_\text{R}^{\top}, o_{3}, \pv_\text{U}^{\top}, \beta]^{\top}.
\end{equation}
The derivation principles of $\mathbf{I}(\etav)$, $\mathbf{I}(\sv)$ and  $\Jm_\mathrm{S}$ can be found for example in~\cite{SISO_RIS_LoS_2021}. %\red{[ref or appendix]}.

The lower bound of parameters of the state vector can be computed by taking the inverse of the \gls{fim} $\mathbf{I}(\sv)$, from which {the lower bound of RIS and user locations can be calculated as $\sqrt{\text{trace}\left(\mathbf{I}^{-1}(\sv)\right)_{[1:3]}}$ and $\sqrt{\text{trace}\left(\mathbf{I}^{-1}(\sv)\right)_{[5:7]}}$, respectively. Similarly, the lower bound of RIS orientation and clock offset can be acquired as $\sqrt{\mathbf{I}^{-1}(\sv)_{[4]}}$ and $\sqrt{\mathbf{I}^{-1}(\sv)_{[8]}}$}. Corresponding numerical examples will be provided in Section IV.

\begin{algorithm}[!t]
\caption{Proposed initialization and snapshot estimation methods}\label{alg:Init_Snapshot}%[H]
\begin{algorithmic}[1]
% \SetAlgoLined
%\Procedure{}{}
\State Based on the direction induced by $\hat{\bm{\varphiv}}_{\text{BU}}$, find the intersection with the area of user $\mathcal{A}_{\text{U}}$. This leads to a distance range of $[d_{\min},d_{\max}]$ between BS and user. %and the direction/line formed by $\hat{\bm{\varphiv}}_{\text{BU}}$ $\longrightarrow$ the \gls{los} distance segment where the user may be located, which we denote as $\bm{d}_{0}$
\State \textbf{For} every candidate LoS distance $\check{d}_{0}\in [d_{\min},d_{\max}]$
\State\indent Estimate the clock offset $\check{\beta}$ as $\check{\beta}=\hat{\tau}_{\text{BU}}-\check{d}_{0}/c$

\State\indent Estimate  the user location  as %\par 
$\check{\pv}_{\text{U}}=\pv_{\text{B}}+\hat\tv_{\mathrm{BU}}\check{d}_{0}$
\State\indent Compute the distance of path (R) $\check{d}_{\text{R}}=(\hat{\tau}_\text{R}-\check{\beta})c$

\State\indent Determine an ellipsoid (denoted by $E_\text{R}$) with \par focal points $\pv_{\text{B}}$ and $\check{\pv}_{\text{U}}$ and focal distance $\check{d}_{\text{R}}$
%on which the RIS is located can be\par established based on $\check{d}_{\text{R}}$ together with $\pv_{\text{B}}$ and $\check{\pv}_{\text{U}}$ as\par the focal points, which we denote as $E_\text{R}$

\State\indent Intersect the ellipsoid $E_\text{R}$ and the line formed \par by  $\hat\tv_{\mathrm{BU}}(\hat{\bm{\varphiv}}_{\text{BR}})$, to determine the RIS location $\check{\pv}_{\text{R}}$

\State\indent \textbf{For} every $\check{o}_{3}$ in the orientation prior $\mathcal{O}_{\text{R}}$

\State\indent\indent Predict the intermediate \gls{aoa} measurements\par{~~~~~} $\check{\vartheta}_{2}$, $\check{\vartheta}_{3}$~\eqref{eq:coupled_AoA} based upon $\check{o}_{3}$, $\check{\pv}_{\text{U}}$, $\check{\pv}_{\text{R}}$ and $\pv_{\text{B}}$

\State\indent\indent Compute the metric: \par  %$\Delta\left(\check{d}_{0},\check{o}_{3}\right)$ as\par{~~~~~} 
 \hspace{1.2cm}$\Delta\left(\check{d}_{0},\check{o}_{3}\right)=\Vert[\check{\vartheta}_{2},\check{\vartheta}_{3}]^{\top}-[\hat{\vartheta}_{2},\hat{\vartheta}_{3}]^{\top} \Vert$

\State\indent \textbf{End}

\State \textbf{End}

\State Determine the RIS orientation $\hat{o}_{3}$ and the LoS\par 
\hspace{-0.7cm}distance $\hat{d}_{0}$ by solving
\begin{equation*}
[\hat{d}_{0},\hat{o}_{3}]=\argmin_{\check{d}_{0},\check{o}_{3}}\bm{\Delta}\left(\check{d}_{0},\check{o}_{3}\right)
\end{equation*}

\State Based $\hat{d}_{0}$, determine $\hat{\pv}_{\text{U}}$, $\hat{\beta}$ and $\hat{\pv}_{\text{R}}$% and $\hat{o}_{3}$.

\State Refine user location %The resulting user location 
$\hat{\pv}_{\text{U}}$, clock offset $\hat{\beta}$, RIS location \linebreak $\hat{\pv}_{\text{R}}$ and orientation $\hat{o}_{3}$ with a Gauss-Newton (GN) method in~\eqref{eq:GN}

%\For{\EndFor
%\EndProcedure
\end{algorithmic}
\end{algorithm}

\begin{figure*}[t] % with subfigures;
\centering
    \begin{subfigure}{0.65\columnwidth}
    \centering
    \includegraphics[width=0.99\columnwidth]{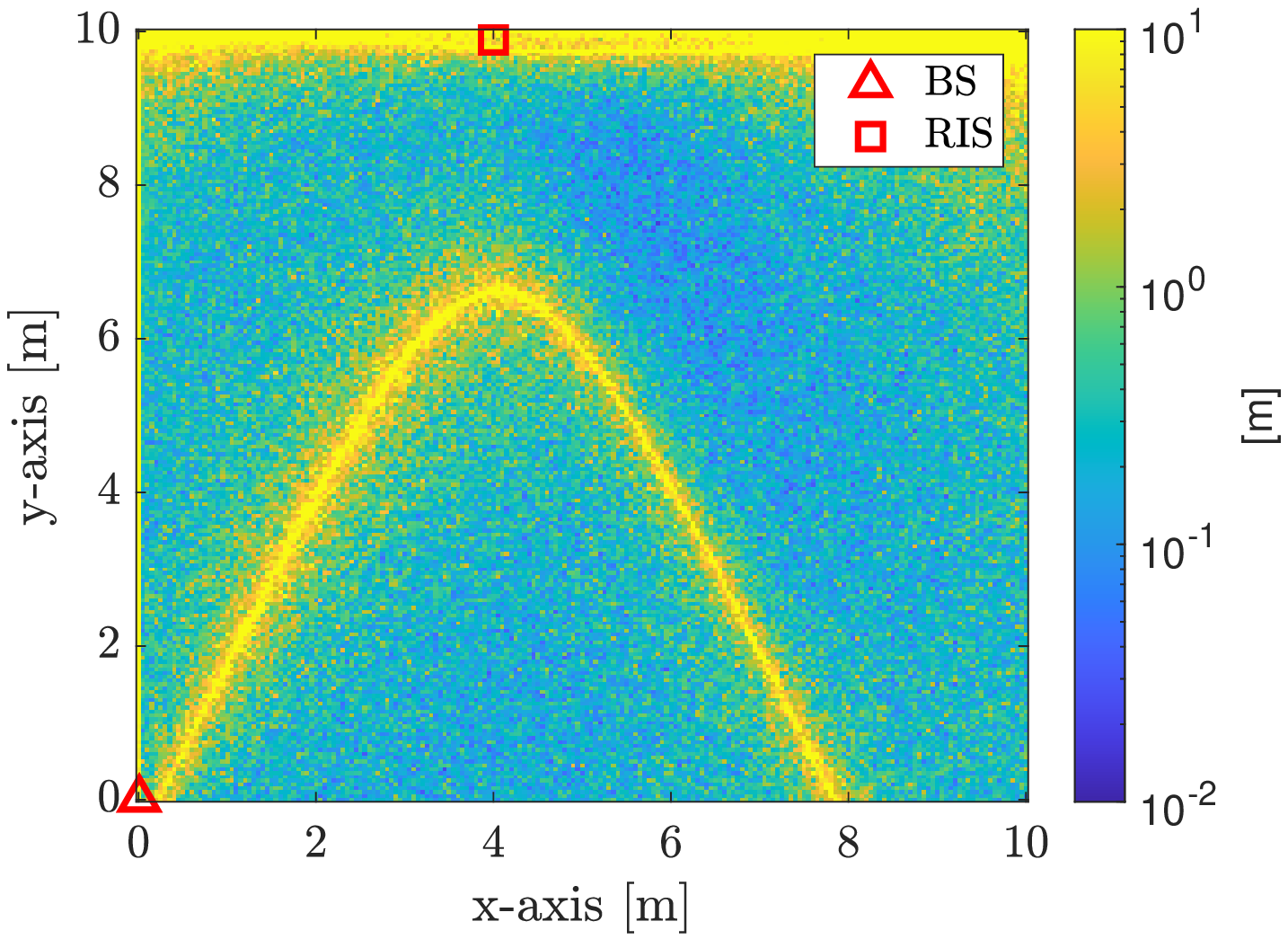}
    \caption{User location.}\label{fig:}
    \end{subfigure}
    \begin{subfigure}{0.65\columnwidth}
    \centering
    \includegraphics[width=0.99\columnwidth]{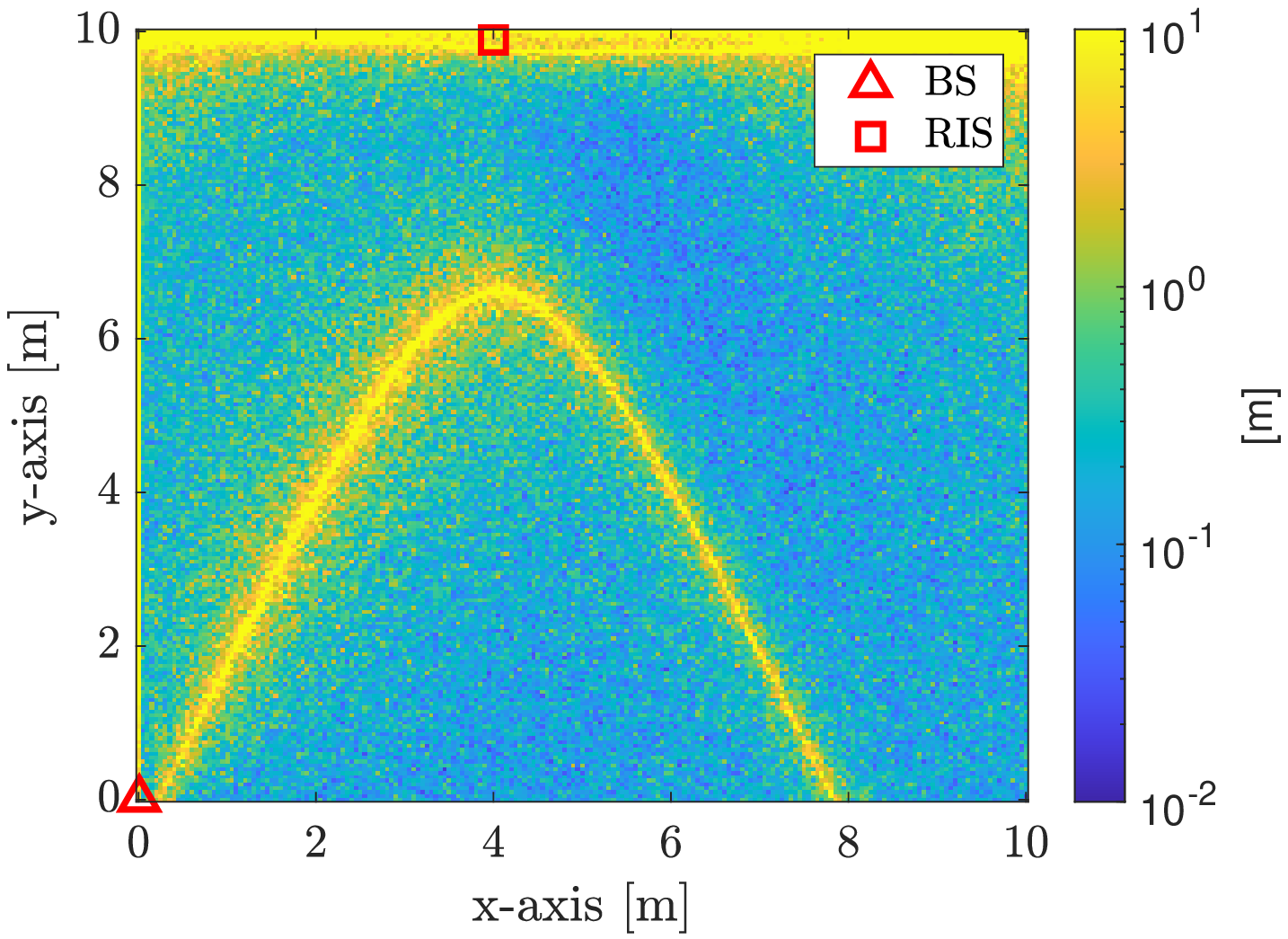}
    \caption{RIS location.}\label{fig:}
    \end{subfigure}
    \begin{subfigure}{0.65\columnwidth}
    \centering
    \includegraphics[width=0.99\columnwidth]{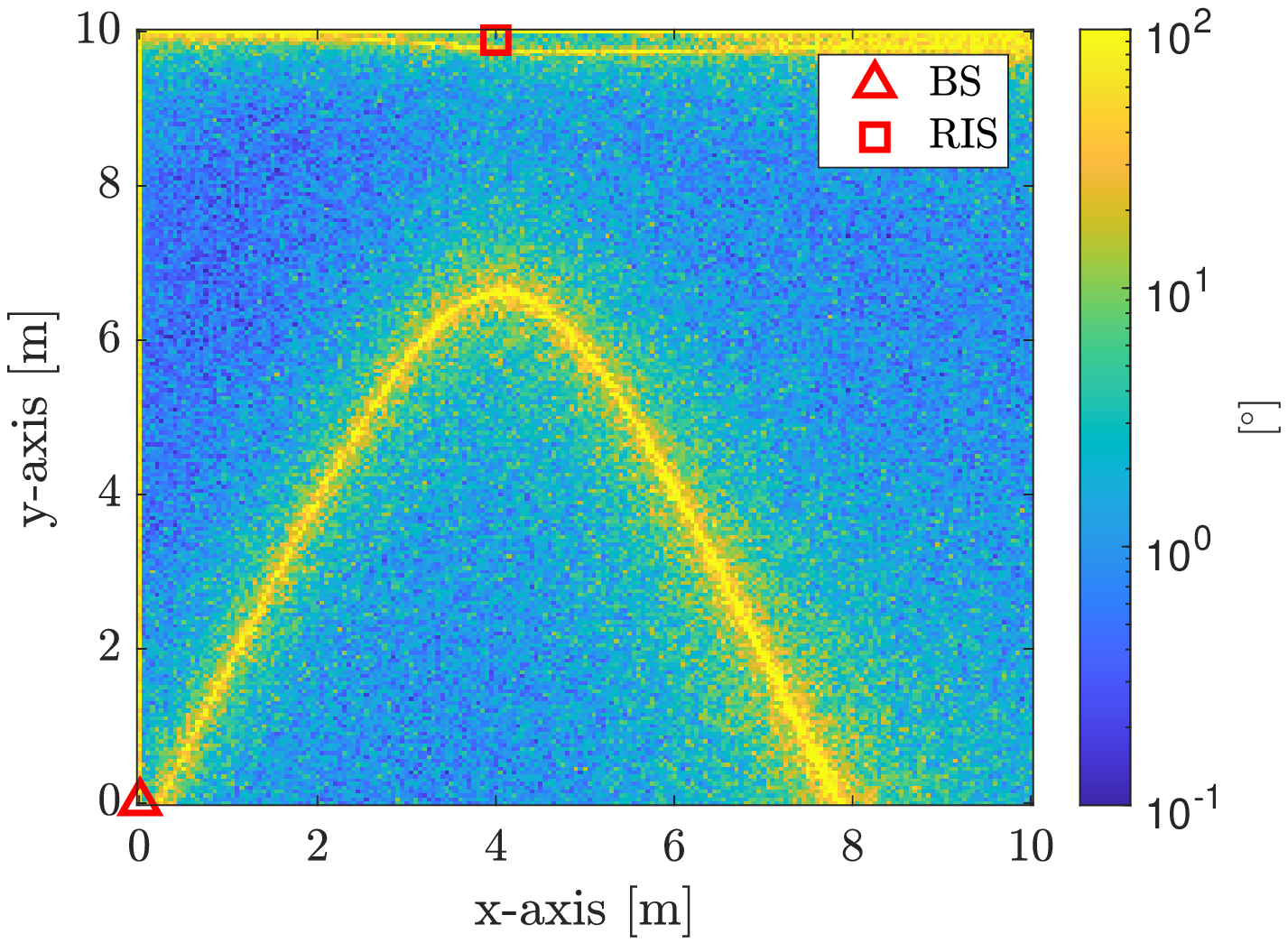}
    \caption{RIS orientation.}\label{fig:}
    \end{subfigure}
    \caption{Visualization of different estimation lower bounds with different user locations: (a) User location error bound, (b) RIS location error bound, (c) RIS orientation error bound. The BS is facing the positive direction of x-axis and RIS is facing the negative direction of y-axis.}
    \label{fig:HeatMap}
\end{figure*}

\subsection{Proposed Initialization and Estimation Methods}

In this subsection, we present an initialization method and a snapshot estimation method for \gls{jrcup} with one or multiple users in the network. For the consistency of notation and formulations, we present first the single-user case, followed by the extension to the  multi-user case. 

Specifically, the step-wise initialization method is outlined from step 1 to step 8 in Algorithm~\ref{alg:Init_Snapshot}, where the outputs are employed as the initial state of user (location and clock offset) and \gls{ris} (location and orientation) before the final estimation. In addition to the channel parameters $\etav_{[1:8]}$ from~\eqref{eq:meas_vec}, the inputs for the initialization contain the prior search area of user, $\mathcal{A}_{\text{U}}$, within the considered environment and that of the RIS orientation $\mathcal{O}_{\text{R}}$. Thereafter, with the searched user and RIS state as the initial guess, the last step of \gls{jrcup} scheme can be carried out using the iterative \gls{gn} method~\cite[Ch. 4]{Sand_2014}, expressed as
\begin{equation}\label{eq:GN}
\hat{\sv}_i=\hat{\sv}_{i-1}+\mathbf{I}^{-1}(\hat{\sv}_{i-1})\Jm_\mathrm{S}^\top \mathbf{\acute{I}}(\etav)\left(\hat{\etav}_{[1:8]}-\etav(\hat{\sv}_{i-1})_{[1:8]}\right),
\end{equation}
in which $i$ denotes the iteration index, while $\mathbf{I}^{-1}(\hat{\sv}_{i-1})$, $\Jm_\mathrm{S}$ and $\mathbf{\acute{I}}(\etav)$ were given in~\eqref{eq:FIM_statePara}. Moreover, the input measurements $\hat{\etav}_{[1:8]}$ are generated using the lower bound of channel parameter, i.e., $\mathbf{\acute{I}}(\etav)$ that approximates a distribution $\mathcal{N}(\etav_{[1:8]}; \hat{\etav}_{[1:8]},(\acute{\textbf{I}}(\etav))^{-1})$. Finally, the predicted measurements $\etav(\hat{\sv}_{i-1})_{[1:8]}$ are evaluated using the estimated state of the $i-1$ iteration. 

In the case of the multi-user scenario, the overall state vector $\sv$ can be redefined as
\begin{equation}
    \sv=\left[\sv_{\text{R}}^{\top},\sv_{1}^{\top},\cdots,\sv_{M}^{\top}\right]^{\top},
\end{equation}
where $\sv_{m}=[\pv_{\text{U},m}^{\top},\beta_{m}]^{\top}$ is the state vector of the $m$th user. The corresponding Jacobian matrix, measurement covariance and measurement vector can all be extended accordingly, and thereon applied in the \gls{gn} algorithm in~\eqref{eq:GN}.

\section{Numerical Results}\label{sec:numerical_results} %  and Analysis Simulation and Performance

% parameter table
\begin{table}[h]
%\begin{table}[t]
\footnotesize 
% \scriptsize
%\small
\centering
\caption{\textsc{Evaluation Assumptions and Numerology}}
\renewcommand{\arraystretch}{1.25}
\begin{tabular} {c | c }
    \hthickline
    \textbf{Parameter} & \textbf{Value}\\
    \hline
    Carrier frequency, $f_c$ & $28$ GHz \\
    %\hline
    Bandwidth, $W$ & $400$ MHz \\
    %\hline
    Number of transmissions, $G$ & 500 \\
    %\hline
    Number of subcarriers, $K$ & 128 \\
    Transmit power & $+30$ dBm \\
    %\hline
    Noise power spectral density, $N_0$ & $-173.8$ dBm/Hz \\
    %\hline
    Noise figure of RX & $10$ dB \\
    BS array size, $N_\text{B}$ & 16~$\times$ 16 \\
    %\hline
    RIS array size, $N_\text{R}$ & 20 $\times$ 20 \\
    %\hline
    Number of RF chains at BS/user & {$  1$}  \\
    Number of GN iterations & 30 \\
    \hthickline
\end{tabular}\label{table:Simulation_parameters}
\normalsize
\vspace{0.3cm}
\end{table}

In this section, we present the achievable accuracy of the proposed \gls{jrcup} scheme in a concrete example scenario at the 28~GHz mmWave band, with the evaluation parameters listed in Table~\ref{table:Simulation_parameters}. In particular, the \gls{bs} is located at $[0,0,0]^\top$ of the applied coordinate system, with antenna array facing the positive of x-axis. The combiner matrix $\wv_{\text{B},g}$ at BS is chosen according to the employed codebooks which we will describe in the next subsection. %differently for each transmission (each element has unit amplitude and random phases). 
The whole area under consideration is 10 m $\times$ 10 m with a RIS located at $[4, 10, 0]^\top$. The user height is set to 5 m below the \gls{bs} to reflect a terrestrial user scenario. {Finally, the distance and orientation intervals are set to 0.1 m and 0.1$^{\circ}$, respectively, which are adopted in both Algorithm~\ref{alg:Init_Snapshot} and the numerical simulations.}

%\vspace{1cm}
\begin{figure}[t] 
\centering
\includegraphics[width=1.0\columnwidth]{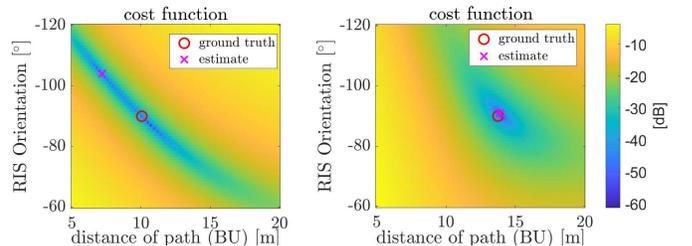}%scale=0.35
\caption{The cost function for user at the blind area (left plot, user located at $[5, 6, -5]$) and non-blind area (right plot, user located at $[9, 8, -5]$) of Fig.~2.}\label{fig:Cost_Func}
\end{figure}

% \red{Higher resolution might be needed, probably add a contour.}

\subsection{Performance Bounds at Different User Locations}
% result figures to be presented;
We start with the heat map of different lower bounds, obtained through the derivations in Section~\ref{eq:FIM}, when user is located throughout the whole map. {In terms of the RIS profile, we assume no prior information is available and apply random RIS coefficients (unit amplitude and random phases) for each transmission as discussed in Section~\ref{subsec:RIS_profile}.} The pattern for clock offsets is approximately the same as that of user location, therefore, is omitted herein for presentation brevity. 

From the obtained pattern in Fig.~\ref{fig:HeatMap}, we see that except for the two blind areas (observed in yellow), both the location and orientation estimations achieve good performance over the considered area as shown in Fig.~\ref{fig:HeatMap}, where the location lower bound is in general under 1 m and the orientation lower bound is under 2$^{\circ}$. In terms of the blind area, one is around the locations that are in line with the RIS or BS array planes, where accurate angle estimates cannot be obtained. Another blind area is the parabola-style zone across the BS and is symmetric to the norm of the RIS array. The reason for such a parabolic blind area lies in the fact that there exist other candidate state vectors that generate the same channel parameters, resulting in ambiguous solutions. 

We further evaluate the reason for the blind area in Fig.~\ref{fig:Cost_Func} from the perspective of the cost function that has been described in Algorithm~\ref{alg:Init_Snapshot}. It is interesting to see that a unique solution exists on the right subplot when the user is located at $[9, 8, -5]$ (at the non-blind area), while ambiguous solutions can be found on the left subplot when the user is located at $[5, 6, -5]$ (at the blind area). This finding indicates that there exists an optimal area for solving the JrCUP problem.
% From the obtained pattern, it is seen that the performance at certain region is clearly better than the performance of other area on the map. In particular, when the user lies in the same plane of the antenna pannel of either RIS or BS, the performance is in general rather poor. This is confirmed by the yellow line regions on the top and on the left of the heat map, when the user is located along the $\pm$ 90 $^{\circ}$ directions with respect to the antenna boresight of RIS or BS. Furthermore, the achievable lower bound is in general under 1 m (the green yellow color) with certain exceptions (blind area) which is demonstrated in all the subplots of Fig.~\ref{fig:HeatMap}. 
% In these blind area, there exist other candidate state vectors that generate the same channel parameters. 

%Show there exist optimal location for RIS calibration. \red{using random RIS profile?}\\
%\textcolor{blue}{1 figure with 3 / 4 subplots, then assume user to be located within a region with good performance. That is the region where the random users are generated for multi-user figure.}\\

\subsection{Impacts of RIS Size and Known States}
\begin{figure}[t!]
\begin{minipage}[b]{0.98\linewidth}
  \centering
    % This file was created by matlab2tikz.
%
%The latest updates can be retrieved from
%  http://www.mathworks.com/matlabcentral/fileexchange/22022-matlab2tikz-matlab2tikz
%where you can also make suggestions and rate matlab2tikz.
%

\begin{tikzpicture}
[scale=1\columnwidth/10cm,font=\normalsize]
\begin{axis}[%
width=8.5cm,
height=3.2cm,
at={(0,0)},
scale only axis,
% xmode=log,
xmin=4,
xmax=900,
xminorticks=true,
xticklabel style = {font=\footnotesize	,yshift=0.5ex},
xlabel style={font=\normalsize\color{white!15!black}, yshift=1 ex},
xlabel={RIS array size $N_\text{R}$},
ymode=log,
ymin=0.1,
ymax=100,
yminorticks=true,
yticklabel style = {font=\footnotesize,xshift=0.5ex},
ylabel style={font=\normalsize\color{white!15!black}, yshift= -1.5 ex},
ylabel={lower bound [m]},
axis background/.style={fill=white},
xmajorgrids,
% xminorgrids,
ymajorgrids,
% yminorgrids,
% legend style={legend cell align=left, align=left, draw=white!15!black}
legend style={font=\footnotesize, at={(.65, 0.55)}, anchor=south west, legend cell align=left, align=left, draw=white!5!black, legend columns=1}
]
\addplot [color=black, line width=1.0pt]
  table[row sep=crcr]{%
4	116.730410174618\\
16	26.3643340691705\\
36	11.8564537119198\\
64	7.19696258550764\\
100	4.65650983593376\\
144	3.23980900317983\\
196	2.55497222700821\\
256	1.97676155264379\\
324	1.60725771503043\\
400	1.3075478561026\\
484	1.17616836093182\\
576	1.04351603982154\\
676	0.927357165638609\\
784	0.904008113639193\\
900	0.752892564902318\\
};
\addlegendentry{Benchmark}

\addplot [color=red, mark=square, mark options={solid, red}, line width=1.0pt]
  table[row sep=crcr]{%
4	41.1603912826144\\
16	10.4427131807877\\
36	4.12862319825998\\
64	2.41689956283348\\
100	1.61589020064922\\
144	1.15350161774988\\
196	0.875987508569506\\
256	0.691906163479664\\
324	0.574833829119358\\
400	0.524299666338739\\
484	0.420291497475204\\
576	0.370693106578634\\
676	0.334692657096129\\
784	0.311442148154807\\
900	0.270277664899129\\
};
\addlegendentry{Known $\mathbf{p}_{\text{R},y}$}

\addplot [color=green, mark = diamond, line width=1.0pt]
  table[row sep=crcr]{%
4	41.9548446299791\\
16	10.8348285164007\\
36	4.76108596616722\\
64	2.72524649411163\\
100	1.90206571373893\\
144	1.36262916939028\\
196	1.017227163355\\
256	0.794506643879914\\
324	0.630966177298254\\
400	0.538284672303847\\
484	0.477928462147334\\
576	0.402630101109552\\
676	0.37068100155914\\
784	0.351692648453257\\
900	0.324745651979661\\
};
\addlegendentry{Known $o_\text{R}$}
\end{axis}

\end{tikzpicture}%
    \vspace{-0.8cm}
    \footnotesize{\centerline{(a) User location lower bound}}\medskip
\end{minipage}
\hfill
\begin{minipage}[b]{0.98\linewidth}
  \centering
    % This file was created by matlab2tikz.
%
%The latest updates can be retrieved from
%  http://www.mathworks.com/matlabcentral/fileexchange/22022-matlab2tikz-matlab2tikz
%where you can also make suggestions and rate matlab2tikz.
%
\definecolor{mycolor1}{rgb}{0.00000,1.00000,1.00000}%
\begin{tikzpicture}
[scale=1\columnwidth/10cm,font=\normalsize]
\begin{axis}[%
width=8.5cm,
height=3.2cm,
at={(0,0)},
scale only axis,
% xmode=log,
xmin=4,
xmax=900,
xminorticks=true,
xticklabel style = {font=\footnotesize,yshift=0.5ex},
xlabel style={font=\normalsize\color{white!15!black}, yshift=1 ex},
xlabel={RIS array size $N_\text{R}$},
ymode=log,
ymin=0.1,
ymax=100,
yminorticks=true,
yticklabel style = {font=\footnotesize, xshift=0.5ex},
ylabel style={font=\normalsize\color{white!15!black}, yshift= -1.5 ex},
ylabel={lower bound  [m]},
axis background/.style={fill=white},
xmajorgrids,
% xminorgrids,
ymajorgrids,
% yminorgrids,
% legend style={legend cell align=left, align=left, draw=white!15!black}
legend style={font=\footnotesize, at={(.65, 0.42)}, anchor=south west, legend cell align=left, align=left, draw=white!5!black, legend columns=1}
]
\addplot [color=black, line width=1.0pt]
  table[row sep=crcr]{%
4	124.818069215574\\
16	27.1667492889093\\
36	11.7444669077474\\
64	6.85541593210144\\
100	4.3440149859046\\
144	2.94426088300312\\
196	2.26036172794736\\
256	1.73223073017173\\
324	1.39014124289637\\
400	1.12294819202184\\
484	0.989513624988244\\
576	0.875356841555182\\
676	0.773390744394716\\
784	0.750276740763705\\
900	0.623686610680172\\
};
\addlegendentry{Benchmark}

\addplot [color=red, mark=square, mark options={solid, red}, line width=1.0pt]
  table[row sep=crcr]{%
4	42.6890066398478\\
16	10.3962579813396\\
36	3.95462200118528\\
64	2.23959729112139\\
100	1.45505247492898\\
144	1.01363053263922\\
196	0.755494378896601\\
256	0.587199382132611\\
324	0.480625221710182\\
400	0.433089807926879\\
484	0.344790361392227\\
576	0.302529931443699\\
676	0.270497465479406\\
784	0.250446629551833\\
900	0.215336214008349\\
};
\addlegendentry{Known $\mathbf{p}_{\text{R},y}$}

\addplot [color=green, mark=diamond, line width=1.0pt]
  table[row sep=crcr]{%
4	54.3872084618392\\
16	13.4616233452952\\
36	5.687210180409\\
64	3.14668744289872\\
100	2.1349561366255\\
144	1.49340132807897\\
196	1.09225350414394\\
256	0.839111788884701\\
324	0.65784138840495\\
400	0.554712575717025\\
484	0.487729209159593\\
576	0.408281156366153\\
676	0.372674990801624\\
784	0.351163655621569\\
900	0.323443992231462\\
};
\addlegendentry{Known $o_\text{R}$}

\addplot [color=blue, mark=o, mark options={solid, blue}, line width=1.0pt]
  table[row sep=crcr]{%
4	23.4760152221374\\
16	5.63663439141747\\
36	2.3381642569575\\
64	1.27485603016106\\
100	0.820483073471675\\
144	0.572187873492086\\
196	0.420185752438858\\
256	0.325827721913037\\
324	0.275322751920621\\
400	0.221624195323649\\
484	0.194105790029558\\
576	0.165576083755262\\
676	0.155348775638863\\
784	0.132367996026714\\
900	0.129844599247681\\
};
\addlegendentry{Known $\mathbf{p}_\text{U}$}
\end{axis}

\end{tikzpicture}%
    \vspace{-0.8cm}
    \footnotesize{\centerline{(b) RIS location lower bound}}\medskip
\end{minipage}
\hfill
\begin{minipage}[b]{0.98\linewidth}
  \centering
    % This file was created by matlab2tikz.
%
%The latest updates can be retrieved from
%  http://www.mathworks.com/matlabcentral/fileexchange/22022-matlab2tikz-matlab2tikz
%where you can also make suggestions and rate matlab2tikz.
%
\definecolor{mycolor1}{rgb}{0.00000,1.00000,1.00000}%
\begin{tikzpicture}
[scale=1\columnwidth/10cm,font=\normalsize]
\begin{axis}[%
width=8.5cm,
height=3.2cm,
at={(0,0)},
scale only axis,
% xmode=log,
xmin=4,
xmax=900,
xminorticks=true,
xticklabel style = {font=\footnotesize,yshift=0.5ex},
xlabel style={font=\normalsize\color{white!15!black}, yshift=1 ex},
xlabel={RIS array size $N_\text{R}$},
ymode=log,
ymin=0.10,
ymax=100,
yminorticks=true,
yticklabel style = {font=\footnotesize, xshift=0.5ex},
ylabel style={font=\normalsize\color{white!15!black}, yshift= -1.5 ex},
ylabel={lower bound [$^{\circ}$]},
axis background/.style={fill=white},
xmajorgrids,
% xminorgrids,
ymajorgrids,
% yminorgrids,
% legend style={legend cell align=left, align=left, draw=white!15!black}
legend style={font=\footnotesize, at={(.65, 0.55)}, anchor=south west, legend cell align=left, align=left, draw=white!5!black, legend columns=1}
]
\addplot [color=black, line width=1.0pt]
  table[row sep=crcr]{%
4	78.6622038630481\\
16	18.5132224858453\\
36	8.61692340916404\\
64	5.48086611959646\\
100	3.55195261269117\\
144	2.56620512545641\\
196	2.14243112282779\\
256	1.61984320727389\\
324	1.35958119087913\\
400	1.12638435589113\\
484	1.05166665349278\\
576	0.908487136859112\\
676	0.809200927792594\\
784	0.77665192505582\\
900	0.649566734268315\\
};
\addlegendentry{Benchmark}

\addplot [color=red, mark=square, mark options={solid, red}, line width=1.0pt]
  table[row sep=crcr]{%
4	50.714736918724\\
16	14.2969511976187\\
36	5.9065227054452\\
64	3.57982463511622\\
100	2.46003160104557\\
144	1.79578588691733\\
196	1.39346574524942\\
256	1.1169417001591\\
324	0.940119514758425\\
400	0.865555629456816\\
484	0.701022887225221\\
576	0.62438912280109\\
676	0.563871114244936\\
784	0.526279061779818\\
900	0.453884732503837\\
};
\addlegendentry{Known $\mathbf{p}_{\text{R},y}$}

% \addplot [color=green, mark=diamond, line width=1.0pt]
%   table[row sep=crcr]{%
% 4	455.58515322021\\
% 16	202.827314330938\\
% 36	97.8802975751134\\
% 64	55.2184604857656\\
% 100	36.6355457414918\\
% 144	26.8574959394748\\
% 196	18.6662348953015\\
% 256	15.8645441411787\\
% 324	13.0911547432234\\
% 400	10.8223997989553\\
% 484	9.84241904416631\\
% 576	8.58924513149715\\
% 676	7.33968527273102\\
% 784	6.93824252358748\\
% 900	6.66753415310282\\
% };
% \addlegendentry{Known $o_\text{R}$}

\addplot [color=blue, mark=o, mark options={solid, blue}, line width=1.0pt]
  table[row sep=crcr]{%
4	27.780148901007\\
16	7.32380916715698\\
36	3.28992467174507\\
64	1.91885294723113\\
100	1.30720117597311\\
144	0.956212801523721\\
196	0.730857737778366\\
256	0.58597201542883\\
324	0.509063814089631\\
400	0.419319464591656\\
484	0.374307087985788\\
576	0.324340200605568\\
676	0.308278691848956\\
784	0.265493980072676\\
900	0.262730040187476\\
};
\addlegendentry{Known $\mathbf{p}_\text{U}$}
\end{axis}

\end{tikzpicture}%
    \vspace{-0.8cm}
    \footnotesize{\centerline{(c) RIS orientation lower bound}}\medskip
\end{minipage}
\caption{Lower bounds as functions of \gls{ris} size $N_{\text{R}}$.}
\label{fig:bound_vs_RIS_size}
\end{figure}
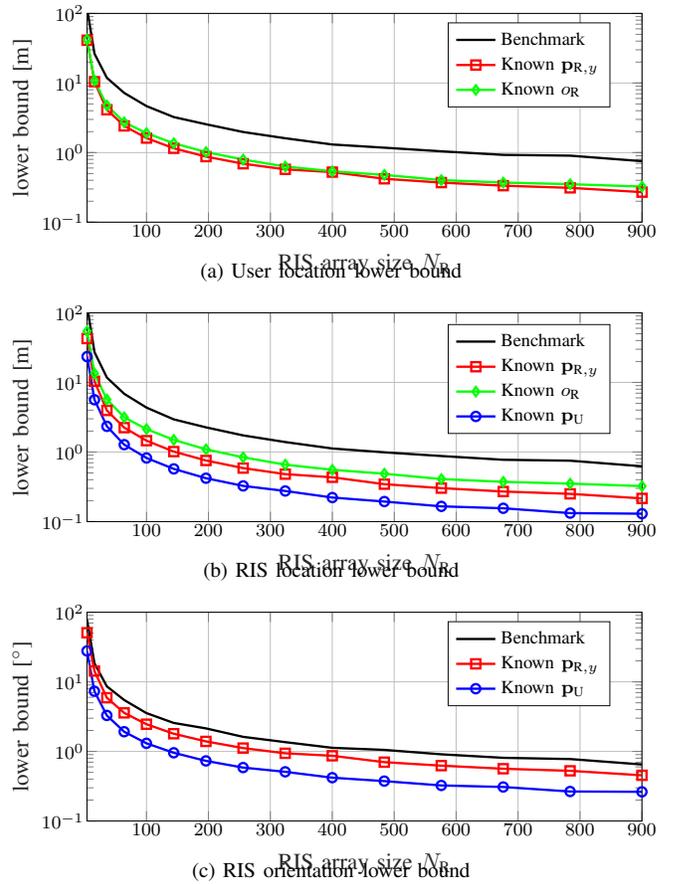

We continue by evaluating the performance bounds for the different state parameters as functions of RIS sizes (i.e., the overall number of RIS elements $N_{\text{R}}$), while also considering different special cases of known states.
The state of \gls{bs} and \gls{ris} remain the same as in Fig.~\ref{fig:HeatMap}, while the user is set at an example location of $[8, 8, -5]^\top$ in the non-blind area. 
The lower bounds for user location, RIS location, and RIS orientation are respectively shown in Fig.~\ref{fig:bound_vs_RIS_size}, in which the benchmark scenario (black curve) assumes the user location, clock offset, RIS location, and RIS orientation are all unknown, yielding overall 8 unknowns. Compared to the benchmark scenario, we find that when one coordinate of RIS location (i.e., $\mathbf{p}_{\text{R},y}$) is known, the performance of all the state parameters is improved as demonstrated by the red curves. In particular, the location estimation performance shown in Fig.~\ref{fig:bound_vs_RIS_size}a and Fig.~\ref{fig:bound_vs_RIS_size}b benefit more from this scenario than the orientation estimates, shown in Fig.~\ref{fig:bound_vs_RIS_size}c. As of the green curve scenario when the RIS orientation is assumed to be accurately known, the achieved performance is on a similar level as the red curve scenario because there are overall 7 unknowns in these two cases. 

Moving next towards the blue curve scenario, when the user acts as a calibration agent within the system, i.e., the user location $\mathbf{p}_\text{U}$ is known, the lower bound of the RIS state is vastly improved. In this case, the location lower bound and orientation lower bound drop to around 0.3~m and 0.5$^{\circ}$ individually with overall 400 RIS elements. It can also be clearly observed that the performance in all different scenarios becomes better as the RIS size $N_{\text{R}}$ increases. This is intuitive since a large RIS size directs the signal power more efficiently and provides a finer angular resolution. These observations suggest that reliable and accurate information on the state of either the user or the RIS as well as a sufficiently large RIS size undoubtedly improve the performance.

%TDMA scheme for multi-user operation.\\
\subsection{Proposed Estimator Performance vs. Bounds } % Multi-user Simulation Results
Finally, we assess and compare the performance of the proposed estimator against the corresponding lower bounds, while also varying the number of the involved users. The obtained results are shown in Fig.~\ref{fig:RMSE_bound_vs_numUser} where the dashed lines with different colors represent the accuracy of initial state, which is calculated using the searching method described in Algorithm~\ref{alg:Init_Snapshot}. Such accuracy can be considered as a reference benchmark. Moreover, the solid lines are the lower bounds of different state parameters, whereas the star markers represent the proposed estimator RMSE over 100 trials. The user locations are randomly generated in each trial with an \gls{ofdma} resource allocation scheme where each user possesses equal bandwidth and the same physical height. Learning from the performance pattern in Fig.~\ref{fig:HeatMap}, we assume that the users are uniformly sampled in a $3\times3$ m$^2$ area $\mathcal{A}_{\text{U}}$ with the start point located at $[6.5, 5.5, -5]^\top$. In such a way, the blind area in Fig.~\ref{fig:HeatMap} where the positioning solution cannot be uniquely identified is avoided. From the numerical results presented in Fig.~\ref{fig:RMSE_bound_vs_numUser}, we can see that with more users in the network, the joint estimation performance improves due to stronger geometric restraint and more information obtained from more measurements. In other words, the stronger geometric constraint formed by more users enhances the accuracy of \gls{ris} state, which in turn helps the user states after the iterations via \gls{gn} approach. It can be observed that the performance of the iterative \gls{gn} method can approach the corresponding analytical lower bound. The proposed method can thus be considered as an efficient estimation solution for \gls{jrcup}.

% \begin{figure}[!t] % single figure
% \centering
% \includegraphics[scale=0.55]{Figures/RMSE_bound_vs_number_of_users_all_States_2Legends.eps} 
% \caption{The RMSE and lower bound as a function of overall number of users.}
% \label{fig:RMSE_bound_vs_numUser}
% \end{figure}

\begin{figure}[t!] % single figure
\centering
% This file was created by matlab2tikz.
%
%The latest updates can be retrieved from
%  http://www.mathworks.com/matlabcentral/fileexchange/22022-matlab2tikz-matlab2tikz
%where you can also make suggestions and rate matlab2tikz.
%
\definecolor{mycolor1}{rgb}{0.00000,0.44700,0.74100}%
\definecolor{mycolor2}{rgb}{0.85000,0.32500,0.09800}%
\definecolor{mycolor3}{rgb}{0.92900,0.69400,0.12500}%
\definecolor{mycolor4}{rgb}{0.49400,0.18400,0.55600}%
\definecolor{mycolor5}{rgb}{0.46600,0.67400,0.18800}%
\definecolor{mycolor6}{rgb}{0.30100,0.74500,0.93300}%
\definecolor{mycolor7}{rgb}{0.63500,0.07800,0.18400}%
\definecolor{mycolor1}{rgb}{0.00000,1.00000,1.00000}%
\definecolor{blue}{rgb}{0.00000,0.00000,1.00000}%
\begin{tikzpicture}
[scale=1\columnwidth/10cm,font=\normalsize]
\begin{axis}[%
width=8.5cm,
height=5cm,
at={(0,0)},
scale only axis,
% xmode=log,
xmin=1,
xmax=10,
xminorticks=true,
xticklabel style = {font=\footnotesize,yshift=0.5ex},
xlabel style={font=\normalsize\color{white!15!black}, yshift=1 ex},
xlabel={number of users},
ymode=log,
ymin=0.10,
ymax=20,
yminorticks=true,
yticklabel style = {font=\footnotesize,xshift=0.5ex},
ylabel style={font=\normalsize\color{white!15!black}, yshift= -1.5 ex},
ylabel={lower bounds [ns]/[m]/[$^{\circ}$]},
axis background/.style={fill=white},
xmajorgrids,
% xminorgrids,
ymajorgrids,
% yminorgrids,
% legend style={legend cell align=left, align=left, draw=white!15!black}
legend style={font=\footnotesize, at={(.01, 0.01)}, anchor=south west, legend cell align=left, align=left, draw=white!5!black, legend columns=2}
]
\addplot [color=mycolor4, line width=1.0pt, only marks, mark=square, mark options={solid, mycolor4}]
  table[row sep=crcr]{%
1	7.76620524782572\\
2	5.52979242381817\\
3	5.35415970323622\\
4	4.96645979738039\\
5	4.87699456308346\\
6	4.8558850975271\\
7	4.55309842236508\\
8	4.51696812586383\\
9	4.50468227175343\\
10	4.19102601905239\\
};
\addlegendentry{clock offset (GN)}
\addplot [color=mycolor3, line width=1.0pt, only marks, mark=o, mark options={solid, mycolor3}]
  table[row sep=crcr]{%
1	2.58256039605346\\
2	1.86520765530742\\
3	1.84398444895554\\
4	1.72615096421709\\
5	1.71073917125072\\
6	1.69871258484352\\
7	1.61003129729408\\
8	1.60967536661768\\
9	1.60085407995423\\
10	1.50026830516502\\
};
\addlegendentry{user location (GN)}

\addplot [color=mycolor2, line width=1.0pt, only marks, mark=diamond, mark options={solid, mycolor2}]
  table[row sep=crcr]{%
1	1.38343961414088\\
2	1.0024556009181\\
3	0.974906537813977\\
4	0.863141048899921\\
5	0.849278124140088\\
6	0.801796384209669\\
7	0.801672383389282\\
8	0.773923721282022\\
9	0.724161132048867\\
10	0.697149560402821\\
};
\addlegendentry{RIS orientation (GN)}

\addplot [color=blue, line width=1.0pt, only marks, mark=asterisk, mark options={solid, blue}]
  table[row sep=crcr]{%
1	0.944870198484814\\
2	0.626156570315268\\
3	0.524137158784946\\
4	0.428259761493596\\
5	0.396957619264871\\
6	0.352689042084876\\
7	0.301658633413792\\
8	0.298280479740552\\
9	0.295932406062623\\
10	0.295744617297525\\
};
\addlegendentry{RIS location (GN)}

\addplot [color=blue, dashed, line width=1.0pt, forget plot]
  table[row sep=crcr]{%
1	1.7\\
2	1.7\\
3	1.7\\
4	1.7\\
5	1.7\\
6	1.7\\
7	1.7\\
8	1.7\\
9	1.7\\
10	1.7\\
};
\addplot [color=mycolor2, dashed, line width=1.0pt, forget plot]
  table[row sep=crcr]{%
1	5.4\\
2	5.4\\
3	5.4\\
4	5.4\\
5	5.4\\
6	5.4\\
7	5.4\\
8	5.4\\
9	5.4\\
10	5.4\\
};
\addplot [color=mycolor3, dashed, line width=1.0pt, forget plot]
  table[row sep=crcr]{%
1	3.4\\
2	3.4\\
3	3.4\\
4	3.4\\
5	3.4\\
6	3.4\\
7	3.4\\
8	3.4\\
9	3.4\\
10	3.4\\
};
\addplot [color=mycolor4, dashed, line width=1.0pt, forget plot]
  table[row sep=crcr]{%
1	11\\
2	11\\
3	11\\
4	11\\
5	11\\
6	11\\
7	11\\
8	11\\
9	11\\
10	11\\
};

\addplot [color=blue, line width=1.0pt, forget plot]
  table[row sep=crcr]{%
1	0.903211356821605\\
2	0.624899017173034\\
3	0.500116470824417\\
4	0.424671357074112\\
5	0.400656939816408\\
6	0.350151145199553\\
7	0.318828348324153\\
8	0.30868684511255\\
9	0.281800805314146\\
10	0.263851355215744\\
};
\addplot [color=mycolor2, line width=1.0pt, forget plot]
  table[row sep=crcr]{%
1	1.29927566266672\\
2	1.03823690633363\\
3	0.92114737928758\\
4	0.851431650882299\\
5	0.803837404486121\\
6	0.768182241622717\\
7	0.740410734703123\\
8	0.71806674963539\\
9	0.699206590348898\\
10	0.684335211127046\\
};
\addplot [color=mycolor3, line width=1.0pt, forget plot]
  table[row sep=crcr]{%
1	2.28776905431779\\
2	1.90745744787027\\
3	1.67732098640219\\
4	1.67123212596134\\
5	1.5565307509314\\
6	1.52861150998775\\
7	1.52045290853746\\
8	1.49321958868658\\
9	1.49117463584724\\
10	1.48281515682372\\
};
\addplot [color=mycolor4, line width=1.0pt, forget plot]
  table[row sep=crcr]{%
1	7.00227537094567\\
2	5.77780397810259\\
3	5.0199365637346\\
4	4.99621604408773\\
5	4.6306010222541\\
6	4.51218506494266\\
7	4.47791522681325\\
8	4.41448867101043\\
9	4.39219675084615\\
10	4.35932277537767\\
};

\addplot [color=black, dashed, line width=1.0pt]
  table[row sep=crcr]{%
0	11\\
};
\addlegendentry{initial accuracy (coarse)}

\addplot [color=black, line width=1.0pt]
  table[row sep=crcr]{%
0	11\\
};
\addlegendentry{lower bounds}
% \addplot [color=green, mark=diamond, line width=1.0pt]
%   table[row sep=crcr]{%
% 4	455.58515322021\\
% 16	202.827314330938\\
% 36	97.8802975751134\\
% 64	55.2184604857656\\
% 100	36.6355457414918\\
% 144	26.8574959394748\\
% 196	18.6662348953015\\
% 256	15.8645441411787\\
% 324	13.0911547432234\\
% 400	10.8223997989553\\
% 484	9.84241904416631\\
% 576	8.58924513149715\\
% 676	7.33968527273102\\
% 784	6.93824252358748\\
% 900	6.66753415310282\\
% };
% \addlegendentry{Known $o_\text{R}$}

% \addplot [color=blue, mark=o, mark options={solid, blue}, line width=1.0pt]
%   table[row sep=crcr]{%
% 4	27.780148901007\\
% 16	7.32380916715698\\
% 36	3.28992467174507\\
% 64	1.91885294723113\\
% 100	1.30720117597311\\
% 144	0.956212801523721\\
% 196	0.730857737778366\\
% 256	0.58597201542883\\
% 324	0.509063814089631\\
% 400	0.419319464591656\\
% 484	0.374307087985788\\
% 576	0.324340200605568\\
% 676	0.308278691848956\\
% 784	0.265493980072676\\
% 900	0.262730040187476\\
% };
% \addlegendentry{Known $\mathbf{p}_\text{U}$}
\end{axis}

\end{tikzpicture}%
\vspace{-0.8cm}
\caption{The estimator RMSE and corresponding lower bounds as functions of the overall number of users.}
\label{fig:RMSE_bound_vs_numUser}
\end{figure}
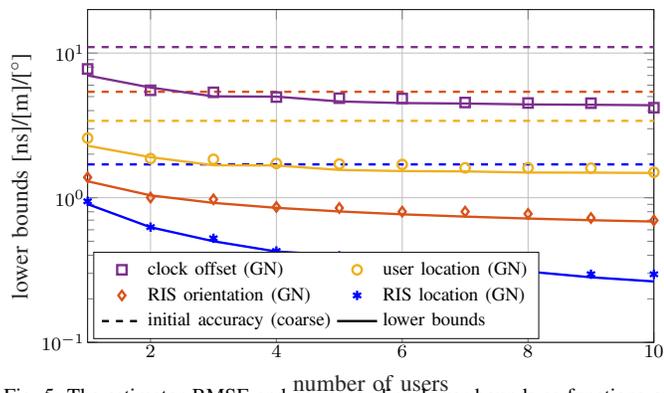

\iffalse

\begin{figure}[t] % single figure
\centering
\includegraphics[scale=0.55]{Figures/RMSE_bound_vs_number_of_users_all_States_2Legends.eps} 
%\include{Figures/fig-5}
%\vspace{-0.8cm}
\caption{The RMSE and lower bound as a function of overall number of users.}
\label{fig:RMSE_bound_vs_numUser}
\end{figure}

\fi

%For the next paper/steps. Fig. 5: Simple beamforming optimization (on RIS, not BS). X-axis: Number of measurement (profile optimized 1 step by 1 step) \red{angle prior accuracy or number of iterations}. Y-axis: Snapshot-based error bound.\\

% Fig. 6: Comparison between network-centric and device-centric.

\section{Conclusion}\label{sec:conclusion}

In this paper, we investigated the problem of joint \gls{ris} calibration and user positioning, called  JrCUP, towards intelligent 6G wireless communication systems. The ultimate objective was to jointly estimate the state parameters of both the users, in terms of clock offsets and 3D positions, and the RIS, in terms of array orientation and 3D position. To this end, we first expressed and computed the lower bound for all the state parameters. We then formulated signal processing methods for state initialization and iterative estimation. Our numerical results showed that the geometric impact can be detrimental and therefore needs extra attention and evaluation in the network planning phase. More importantly, we have found and shown that multi-user scenario in general outperforms the single-user case, demonstrating the potential benefits of deploying multiple users in the system. Moreover, the accuracy of the proposed estimation methods was shown to approach the lower bound, indicating that the proposed methods are efficient, and that the \gls{jrcup} problem is efficiently solvable. Our future research topics include the development of computationally efficient joint state tracking methods with moving users as well as the optimization strategy of RIS phase profiles and BS combiners.
\balance 
\bibliographystyle{IEEEtran}
\bibliography{bib_RIS}

% that's all folks
\end{document}